\documentclass{article}
\usepackage{spconf}
\usepackage[pdftex]{graphicx}
\graphicspath{{plots/pdf/},{plots/schematics/}}
\usepackage[cmex10]{amsmath}
\usepackage[caption=false,font=footnotesize]{subfig}
\usepackage{amssymb}
\usepackage{color}
\usepackage{amsthm}
\usepackage{algorithmic}

\providecommand{\Epsilon}{\mathcal{E}}
\renewcommand{\b}[1]{\ensuremath{\mathbf{#1}}}
\providecommand{\p}[1]{\ensuremath{\mathbf{p}_{#1}}}
\providecommand{\q}[1]{\ensuremath{\mathbf{q}_{#1}}}
\providecommand{\qh}[1]{\ensuremath{\overline{\mathbf{q}}_{#1}}}

\providecommand{\h}[1]{\ensuremath{\mathbf{h}_{#1}}}
\renewcommand{\H}{\ensuremath{\mathbf{H}}}
\newcommand{\norm}[1]{\left\lVert#1\right\rVert}
\newcommand{\Nr}[0]{N_{\text{R}}}
\newcommand{\Nt}[0]{N_{\text{T}}}
\newcommand{\hermitian}[0]{*}

\title{Distributed UAV Placement Optimization for Cooperative line-of-sight MIMO Communications}
\name{Samer Hanna, Han Yan, and Danijela Cabric \thanks{This work was supported in part by the CONIX Research Center, one of six centers in JUMP, a Semiconductor Research Corporation (SRC) program sponsored by DARPA.}}
\address{Electrical and Computer Engineering Department, University of California, Los Angeles\\
samerhanna@ucla.edu, yhaddint@ucla.edu, danijela@ee.ucla.edu}
\begin{document}
\maketitle
\begin{abstract}
Cooperative communication using unmanned aerial vehicles (UAVs) is a promising technology for infrastructureless wireless networks. One of the key challenges in UAV based communications is the backhaul throughput. In this paper, we propose optimization of the UAV swarm positions to achieve a high mulitplexing gain in line-of-sight (LoS) MIMO backhaul. We develop two distributed algorithms to position the UAVs such that each UAV  moves a minimal distance to realize the highest capacity LoS MIMO channel. The first approach uses iterative gradient descent (GD) and the second uses iterative brute force (BF). Simulations show that both algorithms can achieve up to 6 times higher capacity compared to the approach relying on random UAV placement, earlier proposed in the literature.  BF has the advantage of not requiring any location information, while GD is less sensitive to errors in motion. %
\end{abstract}
\section{Introduction}
Unmanned aerial vehicles (UAVs) are envisioned to become an integral component in future wireless networks as they open a myriad of opportunities to enhance communications \cite{7470933,Chandrasekharan:2016jd}. UAVs can be used as aerial base stations (BS) to enhance the performance of cellular network access \cite{8115790}, mobile data aggregators to improve coverage of massive Internet-of-Things (IoT) \cite{8038869}, and network reestablishers in disasters \cite{7589913}. Recently, cooperative communication among UAVs has been proposed. This technique relies on creating a virtual antenna array using a swarm of UAVs to control interference in UAV based cellular access \cite{DBLP:journals/corr/abs-1802-10371,UAV_array,7842113}. A major challenge for the cooperative UAVs swarm array is the high throughput requirement for the backhaul. One approach to boost capacity of a backhaul is to exploit MIMO multiplexing capability of UAV array. However, depending on UAVs' placement,  the MIMO channel could have a very low-rank due to line-of-sight (LoS) propagation.

The designs of linear and planar antenna arrays that maximize MIMO LoS channel capacity were  discussed extensively in the literature \cite{gesbert_ula_2002,haustein_ula_2003, Bohagen_ula_2007,Vucetic_ula_2014,Larsson_ura_2005,Bohagen_ula_ura_2007,Madhow_nonuniform_2009,Vucetic_nonuniform_2016}.    In~\cite{6648625}, capacity was optimized via inter-element spacing of UAVs placed in a uniform linear array. 
However, using conventional antenna array geometries for maximizing UAV based MIMO LoS capacity has the drawback of having stringent localization requirements and does not benefit from the freedom to move UAVs independently.
In \cite{6502954} and \cite{Madhow_random_2013}, the authors propose  randomly placing UAVs as a way to improve the LoS MIMO channel degrees of freedom. The optimal average UAV swarm radius is investigated as a function of wavelength and backhaul link distance. The approach requires more receive antennas than transmit antennas for good performance.

In this work, we propose two distributed algorithms that adapt UAV positions to achieve the highest degree of freedom LoS MIMO channel, while minimizing the distance traveled. One algorithm is based on gradient descent (GD) and the other on brute force (BF) optimization. BF does not require any location information, while GD  has lower traveled distance and robustness against motion errors. We show that starting from a random 3D placement, both algorithms can  achieve the highest possible signal to noise and interference ratio (SINR) and capacity, even if the number of transmit and receive antennas is equal. Compared to using an optimal uniform array placement, both algorithms require less traveled distance to achieve almost the same capacity in the presence of localization errors.

\section{System Model}

We consider a system where a base-station (BS) consisting of  $\Nt$ antennas transmits data to a swarm of $\Nr$ UAVs where each UAV has one antenna\footnote{We assume that this is an antenna in the general sense and can be composed of  a phased array used  to provide directivity to combat the high path loss, and/or to assist in localization. Though we treat it as an idealized antenna and do not discuss beam steering or localization.}, and the combined antenna gain between the BS and each UAV is $G$. 
We assume that $\Nr\geq \Nt$. We  denote the position of the $m$th transmit antenna by $\p{m}\in \mathbb{R}^3$ where $m \in \{1,2, \cdots,\Nt\}$, and that of the $n$th UAV by $\q{n}\in \mathbb{R}^3$ where $n \in \{1,2, \cdots, N_{\text{R}}\}$. We assume that UAVs have been performing some sensing, data aggregation, or receiving data for relaying  prior to the backhauling stage. As a result, the starting positions of the UAVs are assumed to be random.

We consider a typical LoS environment between BS and UAVs. 
The MIMO channel $\mathbf{H} \in \mathbb{C}^{N_{\text{R}}\times N_{\text{T}}}$ between the $m$th transmitter and the $n$th receiver is modeled as 
\begin{equation}
\label{eq:h_element}
\{\mathbf{H}\}_{m,n}\triangleq h_{m,n}=\gamma_{m,n}\exp\left(-j({2\pi}/{\lambda}) \norm{\p{m} - \q{n}}\right)
\end{equation}
where $ \gamma_{m,n} = {\lambda}/({4\pi  \norm{\p{m} - \q{n}}})$ is the path loss coefficient and $\lambda$ is the wavelength. We denote the columns of $\H$ by $\h{1}, \h{2}, \cdots, \h{N_{\text{T}}}$.
In our system, we assume that UAVs are placed far from the BS and move within a relatively small volume compared to the distance from the BS, such that 
$\gamma_{m,n} \approx {\lambda}/({4\pi R})$, 
where $R$ is the distance between  the BS and the UAV swarm.

Assuming the high SINR regime, the maximum channel capacity at a distance $R$ can be achieved when the singular values of $\H$ are equal
, i.e., $\sigma_1 = \sigma_2 = \cdots = \sigma_{N_\text{T}}$ \cite[p.399]{Tse_Wireless_2005}.
This occurs when the channel matrix $\H$ has orthogonal columns. Thus for maximum capacity, the following condition should be satisfied
\begin{equation}
\label{eq:orthogonal_columns}
\h{l}^{\hermitian} \h{k} = 0  \ \ \ \ \forall (l,k) \in \mathcal{K},
\end{equation}
where $()^{\hermitian}$ is the Hermitian transpose,
$\mathcal{K}$ is the set of  $(l,k) \in \{1, \cdots, \Nt\}\times\{1, \cdots, \Nt\}$ such that $l\neq k$.

The   inverse condition number (ICN) of $\H$ is 
\begin{equation}
\Epsilon(\H) = {1}/{\kappa(\H)} = {\sigma_{N_\text{T}}}/{\sigma_1}
\end{equation}
where $\kappa(\H)$ is the condition number of $\H$.
Note that $ 0 \leq \Epsilon(\H) \leq 1$, since 
by definition we have $\sigma_1  \geq \cdots \geq \sigma_{N_\text{T}}$, with $\Epsilon(\H)=1$ iff all the singular values are equal (at the maximum MIMO capacity condition) and $\Epsilon(\H)=0$ when two or more columns of $\H$ are linearly dependent. Hence, we use the ICN as a measure for the orthogonality of $\H$.%

Our goal is to achieve a certain orthogonality requirement, determined by the ICN of $\H$, by optimizing the UAV placement, while trying to reduce the distance traveled by each UAV. This optimization problem can be formulated as 
\begin{equation}
\label{eq:problem_formulation}
\begin{aligned}
& \underset{\q{1},\cdots,\q{N_{\text{R}}}}{\text{minimize}} & & \sum_{n=1}^{N_{\text{R}}} \|\qh{n}-\q{n}\|^2 \\
& \text{subject to} & & \Epsilon(\H) \geq \alpha
\end{aligned}
\end{equation} 
 where $\qh{m}$ is initial position of the $m$th UAV, for some constant $0\leq \alpha\leq 1$, which determines how strict is the requirement for channel orthogonality. 

We assume that the transmitter does not have any channel state information (CSI); during the positioning stage, it only transmits channel estimation pilots, while during the communication stage, it does not perform any precoding. 
We use the single stream SINR as a measure of performance, which can be used to calculate the MIMO capacity  \cite[p.415]{Tse_Wireless_2005}. It is  given by $\text{SINR} = ({|\b{w}^*\h{1}|^2})/({\sum_{m=2}^{N_T}|\b{w}^*\h{m}|^2 + \|\b{w} \|^2 \psi_n^2 })$, where $\b{w}$ is the combining vector, and $\psi_n^2$ is the noise variance.
Two combining methods are considered. The first one is spatial zero-forcing (ZF) \cite[p.413]{Tse_Wireless_2005}  where $\b{w}$ is orthogonal to the range space of $\{\h{2}, \cdots, \h{N_{\text{T}}}\}$.
The second one is naive (NV) matched filtering which uses $\b{w}=\h{1}$ for combining.  For reference, we calculate the single stream matched-filtering (MF) based SINR, which corresponds to the ideal situation of no inter-stream interference \cite[p.414]{Tse_Wireless_2005}. Note that if $\H$ is orthogonal,  all three approaches (ZF, NV, MF) are equivalent.

The main challenge in our proposed UAV algorithms arises from the uncertainties in drone localization and the errors that accompany the movements of UAVs. We assume that each UAV uses its inertial sensors to guide its motion and is able to  localize itself globally with the help of a beacon transmitted by the BS antenna array.
 We will consider two types of errors in our evaluation:  localization errors and actuation errors. The localization error is due to inaccuracies in global positioning. It occurs when the $m$th drone believes it is at location $\q{m}$ but it is actually at location    $\q{m} + \b{n}_{\text{loc}}$. 
As for the actuation error, it arises from the inertial sensors and it occurs when the $m$th drone moves with a motion vector $\b{r}$, but instead of ending up at position $\q{m}+\b{r}$, it ends up at position $\q{m}+\b{r}+\b{n}_{\text{act}}$. In this work, we assume that all $\b{n}_{\text{loc}}$ and $\b{n}_{\text{act}}$ are random Gaussian variables with standard deviations $\sigma_{\text{loc}}$ and $\sigma_{\text{act}}$, respectively. We evaluate the performance of proposed algorithms with respect to localization and actuation error variances.

\section{Distributed UAV Position Optimization}
The problem formulation defined in (\ref{eq:problem_formulation})  aims to find the optimal positions of all $\Nr$ nodes. 
Trying to solve the problem centrally at once and sending the chosen positions to the UAVs  is not practical. 
It would require perfect knowledge of the positions of  UAVs and the ability to accurately position them anywhere in space.  However, in practice, the inherent accuracy limit for any localization system and the variability in the actuation of the drones make positioning them exactly in a certain location difficult. Any practical deployment would need to account and adapt to these disturbances in real-time.  Solving optimization centrally on one UAV would be too complex, while off-loading it to the BS would incur a large communication overhead.

Due to these limitations, we recast the problem into a distributed formulation which is more practical. We consider the case where each UAV with some information from the rest of the UAVs tries to optimize its position. By performing several iterations over all the UAVs, the distributed algorithm solves the global problem.

\vspace{-1mm}
\subsection{Problem Formulation}
\vspace{-1mm}
From the perspective of the $m$th UAV at the  $i$th iteration,  the problem can be formulated as follows
\begin{equation}
\begin{aligned}
\label{eq:distributed_formulation}
\underset{\b{r}_m^{(i)}}{\text{minimize} }\ & &   f(\q{m}^{(i)}+\b{r}_m^{(i)}) 
\end{aligned} 
\end{equation}
where  $\q{m}^{(i)}$ is the position of the $m$th UAV at the  $i$th iteration, $\b{r}_m$ is a vector representing the motion of  the UAV  in space, such that  $\q{m}^{(i+1)} = \q{m}^{(i)} + \b{r}_m^{(i)}$ and the objective function is
\begin{equation} 
f( \q{m} ) = \sum_{(l,k)\in \mathcal{K}} |\h{l}^{\hermitian}(\q{m})\h{k}(\q{m})|^2 
\end{equation}
This is equivalent to finding the step which orthogonalizes the channel. The ultimate goal is to have the objective of the distributed optimization problem defined in (\ref{eq:distributed_formulation}) to be close to 0. But, since the change of location of one UAV affects an entire row of the channel matrix according to (\ref{eq:h_element}), a single UAV can not orthogonalize the columns of the matrix on its own.  All UAVs have to move to accomplish this goal. Additionally, the motion of one UAV affects all the others, thus the algorithm needs to iterates over all UAVs several times.
We propose two algorithms to solve the global problem given  in (\ref{eq:problem_formulation}) by iteratively addressing the distributed problem in (\ref{eq:distributed_formulation}). 

\subsection{Gradient descent (GD) location optimization}
\label{subsec:heuristics}
We use gradient descent to find a practical solution to the distributed problem.  
We calculate the gradient of the objective function with respect to $\q{m}$ given by 
\begin{multline*}
\nabla f(\q{m})  = \sum_{(l,k)\in \mathcal{K}} 
\frac{4\pi}{\lambda}(\Re\left\{ \h{-m,l}^{\hermitian}\h{-m,k}\right\}  \times\sin(a)\\-\Im\left\{ \h{-m,l}^{\hermitian}\h{-m,k}\right\} \times\cos(a))\times\left(u(\p{l}-\q{m})-u(\p{k}-\q{m})\right)
\end{multline*}
where $a=\frac{2\pi}{\lambda} (\lVert \p{l}-\q{m}\rVert-\lVert \p{k}-\q{m}\rVert)$, $u(\b{v})$ is a unit vector in direction of $\b{v}$, and $ \h{-m,l}$ is column $l$ except element $m$. The value of $\b{r}_m$ is then calculated using
$\b{r}^{(i)}_m =  - b_{\text{GD}}^{(i)} \nabla f(\q{m}^{(i)})$
where $ b_{\text{GD}}^{(i)}$ is the step size which is a parameter of the algorithm and $b_{\text{GD}}^{(i+1)}= d\ b_{\text{GD}}^{(i)}$, where $d<1$ is a decay factor which reduces the step size every iteration.

In our proposed approach, each UAV decides on its  motion alone, while one  master UAV coordinates between them.  The proposed GD algorithm proceeds as follows. %
\begin{algorithmic}[1]
	\STATE  estimateBroadcastCSIAll()  \{Master, Drones\}
	\FOR {$i = 1$ to \#Iterations  \{Master\}}
	 \STATE estimatePositionAll() \{Master, Drones\}
	\FORALL {$d$ in Drones \{Master\}}
	\STATE  calcGradientAndMoveDrone($d$)  \{Drone $d$\}
	\STATE  estimateBroadcastCSIDrone($d$)  \{Drone $d$\}
	\ENDFOR
	\STATE c=evalChannelOrthogonality()  \{Master\}
	\IF {meetsCriterion(c)  \{Master\}}  
	\STATE exitStartCommunicationStage()  \{Master\}
	\ENDIF
	\ENDFOR
\end{algorithmic}
Note that this algorithm requires position estimation, which adds overhead to the system. 
\subsection{Brute force (BF) location optimization}
The brute force algorithm relies only on channel information to guide UAVs and does not need any position information. 
The idea behind it is simple. Each drone tries a set of positions and determines the one that minimizes the objective function. It can be formally described as moving $\b{r}_m^{(i)} = b_{\text{BF}}^{(i)} \hat{z}^{(i)}$, where 
$\hat{\b{z}} = \underset{\b{z} \in \mathcal{Z}  } {\text{argmin}}f(\q{m} + \b{r}_m^{(i)} \b{z})$ is calculated by evaluating the function  $f(\q{m} + \b{r}_m^{(i)} \b{z})$ for all $\b{z} \in \mathcal{Z}$ and choosing the minimum. Here $\mathcal{Z} = \{\b{0},\pm \b{e}_1,\pm \b{e}_2,\pm \b{e}_3\}$,  $\b{e}_i \in \mathbb{R}^3$ and has one at position $i$ and zero elsewhere.
The proposed BF algorithm proceeds as follows.%
\begin{algorithmic}[1]
	\STATE estimateBroadcastCSIAll() \{Master, Drones\}
	\FOR {$i = 1$ to \#Iterations  \{Master\}}
	\FORALL {$d$ in Drones  \{Master\}}
	\FORALL {$\b{z} \in \mathcal{Z}$ \{Drone $d$\}}
	\STATE moveDrone($d$,$\b{z}$) \{Drone $d$\}
	\STATE estimateCSIandEvalObjectiveDrone($d$) \{Drn $d$\}
	\STATE moveDrone($d$,$-\b{z}$) \{Drone $d$\}
	\ENDFOR
	\STATE $\hat{z}$ = findMinObj() ; moveDrone($d$,$\b{\hat{z}}$) \{Drone $d$\}
	\STATE broadcastCSIDrone($d$) \{Drone $d$\}
	\ENDFOR
	\STATE c=evalChannelOrthogonality()  \{Master\}
	\IF {meetsCriterion(c)  \{Master\}}
	\STATE exitStartCommunicationStage()  \{Master\}
	\ENDIF
	\ENDFOR
\end{algorithmic}

\section{Evaluation}
\begin{figure}[t!]
	\centering
	\subfloat[ICN]{\includegraphics{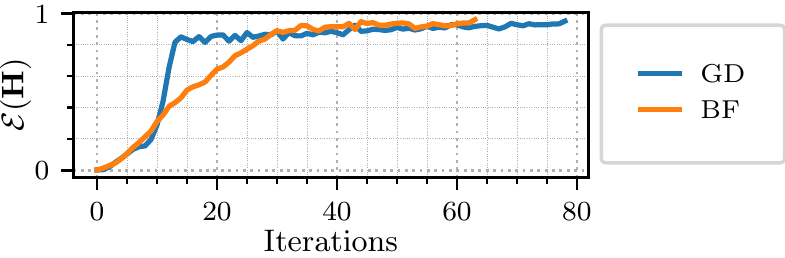}\label{fig:runs_icn}}\\
	\vspace{-3mm}
	\subfloat[SINR]{\includegraphics{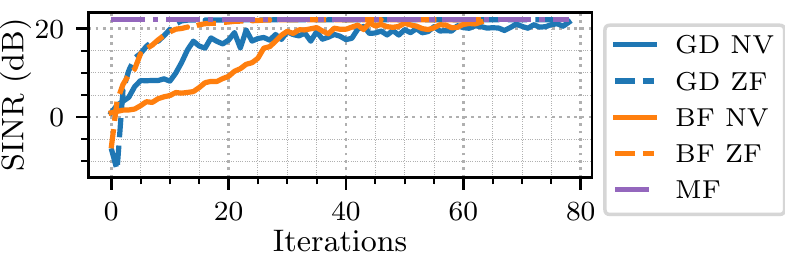}\label{fig:runs_sinr}}
	\vspace{-3mm}
	\caption{ICN and SINR shown against the number of iterations.}
	\label{fig:runs}
	\vspace{-1mm}
\end{figure}

\begin{figure}[t!]
	\centering
	\subfloat[Average Distance per UAV. In solid  ICN 0.5, and in dotted ICN 0.95.]{\includegraphics{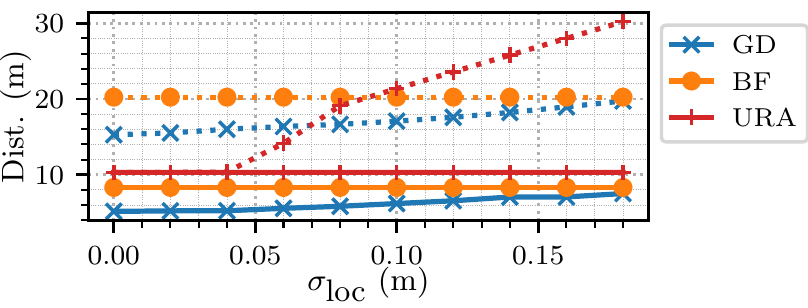}\label{fig:localization_dist}}\\
	\vspace{-3mm}
	\subfloat[SINR. In solid  NV, in dotted ZF SINR, and in dash-dot purple MF]{\includegraphics{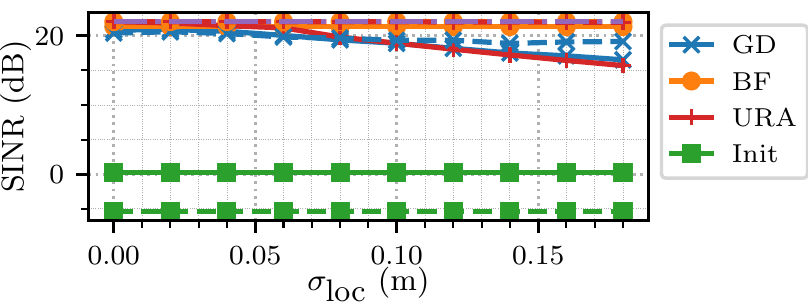}\label{fig:localization_sinr}}
	\vspace{-3mm}
	\caption{Effect of localization error.}
	\label{fig:localization_res}
	\vspace{-1mm}
\end{figure}

\begin{figure}[t!]
	\centering
\includegraphics{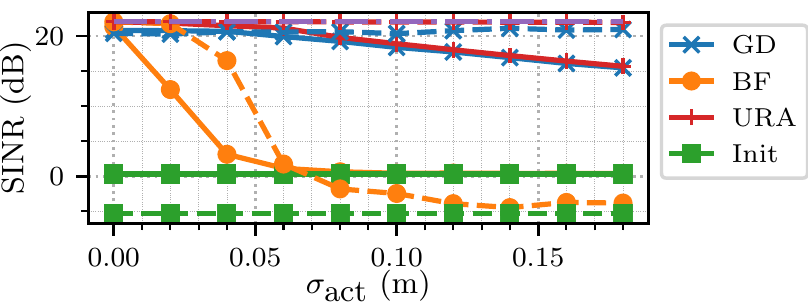}\label{fig:actuation_sinr}
\vspace{-3mm}
	\caption{Effect of actuation error on SINR. ZF SINR is shown in dotted, NV in solid,  and MF in dash-dot purple.}	
	\label{fig:actuation_res}
	\vspace{-1mm}
\end{figure}

The setup used for evaluation is as follows. The BS transmitter consists of 16 antennas arranged in a $4\times 4$ uniform square array, with separation between elements of 25 cm. There are 16 UAVs acting as receivers. At the beginning, UAVs are randomly distributed in a cube of side 50m, whose center is placed 1km away from the BS. The frequency of operation is assumed to be 60 GHz, though the proposed algorithms are valid for any other frequency.  The SNR is assumed to be equal to 10 dB after considering the antenna gains.
BF was tested with a step size of $b_{\text{BF}}=0.3$  while  GD used a step size of $b_{\text{GD}}=0.05$, both with decays of $d=0.999$.

We start by showing the ability of the algorithms to converge to an orthogonal channel. Figure \ref{fig:runs_icn} shows the  ICN obtained by each of the proposed algorithms as a function of number of iterations. Both algorithms ran until achieving an inverse condition number of 0.95. 
  Figure  \ref{fig:runs_sinr}, shows the NV SINR, ZF SINR, and MF in solid, dashed, and dash-dot respectively. We observe that the ZF SINR converges faster and approaches the maximum SINR starting from ICN=0.5. However, when both ZF and NV algorithms converge to ICN larger than 0.95, they have almost identical performance, which advocates for NV combining being a simpler option. 

 In addition to the proposed algorithms, we evaluated a baseline algorithm which consists of directing UAVs to form an optimal uniform rectangular array  (URA) \cite{Bohagen_ula_ura_2007}. Each UAV is directed to a position in the URA that minimizes the total distance traveled by all UAVs.
This algorithm achieves an orthogonal $\H$  at the first iteration if there are no external disturbances, otherwise the UAVs  keep attempting to get to the assigned positions in subsequent iterations. All these methods were tested for different standard deviations of localization  and actuation errors. The curves correspond to the average of 500 random starting position.

Figure \ref{fig:localization_res} shows the performance of the proposed methods against the standard deviation of localization errors.  First, we notice that the BF has the advantage of not requiring any location information, hence its curves remain constant.  The  average distance traveled until ICN 0.5 and 0.95 are reached is shown in solid and dotted lines respectively in Figure \ref{fig:localization_dist}.  We see that GD requires less distance traveled than URA and BF to achieve an ICN of 0.5. The URA method has the advantage of starting with a higher ICN, since it directs the UAVs to a set of positions which orthogonalize the channel. But due to its complete reliance on localization information, as the standard deviation of localization error increases, it starts requiring more distance to achieve a high ICN than both methods. 
In Figure \ref{fig:localization_sinr}, we show the SINR obtained at the end of each of algorithm.  We see that any location optimization gives at least  15 dB improvement in  SINR over random placement even when localization errors are considered. We also notice that both BF and GD give results comparable to URA positioning at a less distance traveled under high localization errors.

The effect of actuation error on the obtained SINR is shown in Figure \ref{fig:actuation_res}. We can see that the BF algorithm is the most affected by the actuation error. This due to the fact it relies on performing 6 back and forth motions given by $\mathcal{Z}$ for each UAV per each iteration step.

\section{Conclusion}
\label{sec:conc}
We proposed two distributed algorithms that adaptively optimize the positions of UAVs in order to improve the LoS MIMO channel capacity. The gradient descent based algorithm is able to converge faster than brute force based algorithm but it requires position information. On the other hand, the brute force algorithm uses only channel information to find an optimized set of positions. Both algorithms are more resilient to localization errors compared to placing the UAVs in a URA as indicated by the required traveled  distance. Gradient descent was shown to be insensitive to actuation errors in contrast to brute force bases algorithm. They were also shown to achieve up to 20 dB SINR improvement compared to the approach of relying on the random placement of the UAVs for a $16\times16$ LoS MIMO channel. 
\clearpage
\bibliographystyle{IEEEbib}
\bibliography{references}

\end{document}